# Toward a systems-level view of mitotic checkpoints

Bashar Ibrahim

Bio System Analysis Group, Friedrich-Schiller-University Jena, and Jena Centre for Bioinformatics (JCB), 07743 Jena, Germany.

E-Mail: bashar.ibrahim@uni-jena.de; Tel.: +49-3641-9-46460; Fax +49-3641-9-46302.

**Summary**

Reproduction and natural selection are the key elements of life. In order to reproduce, the genetic material must be doubled, separated and placed into two new daughter cells, each containing a complete set of chromosomes and organelles. In mitosis, transition from one process to the next is guided by intricate surveillance mechanisms, known as the mitotic checkpoints. Dis-regulation of cell division through checkpoint malfunction can lead to developmental defects and contribute to the development or progression of tumors.

This review approaches two important mitotic checkpoints, the spindle assembly checkpoint (SAC) and the spindle position checkpoint (SPOC). The highly conserved spindle assembly checkpoint (SAC) controls the onset of anaphase by preventing premature segregation of the sister chromatids of the duplicated genome, to the spindle poles. In contrast, the spindle position checkpoint (SPOC), in the budding yeast S. cerevisiae, ensures that during asymmetric cell division mitotic exit does not occur until the spindle is properly aligned with the cell polarity axis. Although there are no known homologs, there is indication that functionally similar checkpoints exist also in animal cells.

# 1. Introduction

Correct DNA segregation during mitosis is a fundamental process that ensures the faithful inheritance of genomic information for the propagation of cell life. Segregation (Figure 1) failures underlie many human health problems, most notably aneuploidy and cancer (Holland and Cleveland, 2009; Suijkerbuijk and Kops, 2008). In order to avoid this catastrophe, cells contain active segregation machineries that grant proper DNA distribution into the daughter cells. This requires binding of proteins ("kinetochore proteins") to DNA sites ("centromeres") and actively transporting the DNA through the cellular space to the new location. When the genome is distributed over several or many chromosomes, each daughter cell must obtain the complete set of chromosomes. This requires a concerted action of combined transport requiring precise regulation ("mitotic checkpoints"). Hence, DNA segregation is an essential, highly ordered process that depends on the assembly and multi-functionality of numerous protein complexes that are regulated both in time and space.

Experimentalists have provided a wealth of information about the components of DNA segregation. However, DNA segregation is inherently complex and highly cross linked. It cannot be understood from reactions on the individual components (proteins, complexes etc) alone, but should be understood through considerations involving many components at the same time. The current experimental techniques are not sufficient to make quantitative predictions. Hence, the integration of experimental and computational approaches is employed for the understanding of biological systems. This approach recently received mainstream attention by scientists following the so called "Systems Biology" approach. Some refer to systems biology as an emergent field that "studies cells as spatiotemporal networks of interacting molecules using an integrative approach of theory (mathematics, physics, engineering), experimental biology (genetics, molecular biology, physiology), and quantitative network-wide analytical measurement (analytical biochemistry, imaging) (Bruggeman, 2007; Kitano, 2002; Klipp et al., 2008)." We may also define systems biology as a tool for understanding molecular biological phenomena (see Figure 2 and 3).

## 2. DNA segregation

Faithful DNA segregation requires three subsystems:

- Kinetochore assembly: the inner and outer kinetochore

- Kinetochore attachment, the spindle assembly checkpoint

- Placing the correct DNA into the right cell, the spindle position checkpoint

### 2.1. Kinetochore assembly: the inner and outer kinetochore

The kinetochore is a multi-protein complex that assembles solely at the centromere of each sister chromatid and contains over 100 proteins (Perpelescu and Fukagawa, 2011). These proteins can be classified into two functional groups: the inner-kinetochore, which is tightly associated with the centromere DNA, and the outer-kinetochore, which interacts with microtubules.

The inner kinetochore is composed of a centromeric CenpA and 16 CCAN (constitutive centromere-associated network) proteins (CenpC, CenpH, CenpI, CenpK to CenpU, CenpW, CenpX) (Okada et al., 2006). The inner kinetochore is relatively stable and present during most of the cell cycle (Black and Cleveland, 2011; Dalal and Bui, 2010; Perpelescu and Fukagawa, 2011), while the outer-kinetochore is thought to be structurally unstable and formed in early mitosis (Cheeseman and Desai, 2008; Maiato et al., 2004).

The outer kinetochore proteins and complexes include the KNL-complex (Spc105/Knl1), Mis12-complex (Dsn1, Nnf1, Nsl1 and Mis12) and the Ndc80-complex (Ndc80/Hec1, Nuf2, Spc24 and Spc25), which are known as the KMN network (Perpelescu and Fukagawa, 2011).

Additionally, the Ska-complex (Ska1, Ska2 and Ska3/Rama1) is essential for the attachment of the chromosome to the spindle microtubules (Jeyaprakash et al., 2012).

***Why use systems biology to study the kinetochore structure?***

Studying the three dimensional structure of the kinetochore is challenging both experimentally and theoretically. It is experimentally difficult because the average diameter of a CCAN protein is 40 Å and connections between them are not visible through microscope techniques. Also, theoretical analysis methods are hindered by the combinatorial explosion of the amount of intermediate complexes and protein assembly states (Tschernyschkow et al., 2013). Conventional or classical modeling approaches based on explicit representations of all intermediate complexes, such as differential equations, cannot account for that combinatorial explosion. A recent S-Phase inner-kinetochore model by (Tschernyschkow et al., 2013) overcomes combinatorial complexity using an implicit representation, combining a rule-based language with a novel particle based simulation approach (Gruenert et al., 2010; Ibrahim et al., 2013). Despite that the method suites the combinatorial complexity, the spatial simulation SRSim is at the very early stage and needs to be faster and able to consider large scale simulations.

## 2.2. Kinetochore attachment, the Spindle Assembly Checkpoint (SAC)

Eukaryotic cells have evolved a conserved surveillance control mechanism for DNA segregation, known as the Spindle Assembly Checkpoint (SAC; (Minshull et al., 1994)). It delays the onset of anaphase until all chromosomes have made amphitelic tight bipolar attachments to the mitotic spindle. Even one misaligned chromosome is sufficient to keep the checkpoint active, yet the mechanism by which this is achieved is still elusive. It is thought that unattached or misaligned kinetochores catalyze the formation of a "wait-anaphase" signal which then diffuses to counter the activation of the ubiquitin ligase anaphase promoting complex/cyclosome (APC) by its coactivator Cdc20 (Cell Division Cycle 20 homolog (Hartwell et al., 1970)). Activation of APC by Cdc20 triggers chromosome segregation by ubiquitination of securin and cyclin B (King et al., 1995; Sethi et al., 1991; Shirayama et al., 1998; Sudakin et al., 1995) (for review see (Robbins and Cross, 2010)). Dysfunction of the SAC leads to aneuploidy (Cimini and Degrassi, 2005; Suijkerbuijk and Kops, 2008) and its reliable function is important for tumor suppression (Holland and Cleveland, 2009; Li et al., 2009).

The core proteins involved in SAC are conserved in all eukaryotes include MAD ("Mitotic Arrest Deficient"; Mad1, Mad2, and Mad3 (in humans: BubR1)) (Li and Murray, 1991) and BUB ("Budding Uninhibited by Benzimidazole"; Bub1, and Bub3) (Hoyt et al., 1991), all of which are conserved among eukaryotes. These proteins work to regulate APC activity and its co-activator, Cdc20.  In addition to these core proteins, the SAC also involves several other components that also are involved in essential aspects of this mechanism. Among these additional components are Aurora-B (Vagnarelli and Earnshaw, 2004) and the "Multipolar spindle-1" protein (Mps1) (Fisk et al., 2004). These components are required for SAC signal amplification and MCC formation. Moreover, several other proteins have been identified in higher eukaryotes that are also involved in carrying out essential aspects of the SAC mechanism e.g., localization or biochemical signaling, for example, the RZZ complex (Karess, 2005; Lu et al., 2009) which composed of Rough Deal"(Rod) (Buffin et al., 2007; Raff et al., 2002) (Zeste White 10) (Raff et al., 2002; Saffery et al., 2000a; Saffery et al., 2000b), (Williams et al., 2003), Zwint-1 (Kops et al., 2005).

The biochemical reactions of SAC activation and maintenance mechanism (Figure 4) can be divided into kinetochore dependent (modules I, III and IV in Figure 4) and cytosolic (kinetochore independent) parts (modules II-VI in Figure 4). These can be thought of as the following modules:

### 2.2.1. Mad2-activation and its function in sequestering Cdc20

Mad2-activation at the kinetochores is commonly seen as the central part of the SAC mechanism. It is known as the "Template model" According to this model (module I in Figure 4), Mad2 in its open conformation (O-Mad2) is recruited to unattached kinetochores by Mad1-bound Mad2 in its closed conformation (C-Mad2) to form the ternary complex Mad1:C-Mad2:O-Mad2* (De Antoni et al., 2005). In this complex O-Mad2* is the "activated" Mad2, i.e., it is stabilized in a conformation which can interact with Cdc20 to form Cdc20:C-Mad2. The kinetic rate coefficients for this interactions have been determined *in vitro* by Simonetta et al. (2009).

### 2.2.2. Autocatalytic amplification of Cdc20:C-Mad2 formation

In addition to its activation via kinetochore-bound Mad1:C-Mad2, O-Mad2 can likewise be activated by Cdc:C-Mad2 (module II in Figure 4) to autocatalytically increase Cdc20:C-Mad2 formation rate *in vitro* (Simonetta et al., 2009). However, the contribution of this autocatalytic loop is minor with the kinetic data given in (Simonetta et al., 2009). The presence of a highly contributing autocatalytic loop would also counteract checkpoint deactivation and is therefore not desirable for live cells from a theoretical point of view (Ibrahim et al., 2008b).

### 2.2.3. MCC formation

Although Cdc20:C-Mad2 can bind to and inhibit APC directly, its inhibitory potency increases greatly in synergy with the Bub3:BubR1 complex (Musacchio and Salmon, 2007). It was shown that Cdc20:C-Mad2 together with Bub3:BubR1 forms the tetrameric mitotic checkpoint complex (MCC, cf. module III in Figure 4), which is a potent inhibitor of APC (Sudakin et al., 2001). The trimeric complex Bub3:BubR1:Cdc20 alone is a potent inhibitor of APC, too (Malureanu et al., 2009). However, the rate of its uncatalyzed formation in the cytosol (module III in Figure 4) is slow (Sudakin et al., 2001). The formation of Bub3:BubR1:Cdc20 is accelerated in the presence of unattached chromosomes (Kulukian et al., 2009) and it may be that MCC forms as an intermediate complex from which O-Mad2 rapidly dissociates (Kulukian et al., 2009; Malureanu et al., 2009; Medema, 2009).

### 2.2.4. APC inhibition

The APC can be inhibited in multiple ways, and complexes of APC together with either Cdc20:C-Mad2, Bub3:BubR1:Cdc20, MCC or MCF2 have been found to be inactive (Eytan et al., 2008; Herzog et al., 2009; Kulukian et al., 2009; Malureanu et al., 2009; Medema, 2009; Sudakin et al., 2001). However, the demand mechanisms for binding the inhibitory complexes to the APC are the subject of current research.

Despite its possibly transient nature in Bub3:BubR1:Cdc20 formation, MCC has been found stably bound to APC in mitosis (Herzog et al., 2009). The MCC may form more stably at unattached kinetochores and recruit APC from the cytoplasm (module IV in Figure 4). The MCC sub-complexes Bub3:BubR1:Cdc20 and Cdc20:C-Mad2 can bind to the APC independently from unattached kinetochores (module V in Figure 4) although their binding may be facilitated in a kinetochore-dependent manner (Kulukian et al., 2009; Malureanu et al., 2009; Medema, 2009). The MCC may also bind to APC in a kinetochore independent manner to eventually inhibit APC by releasing O-Mad2, forming the stable Bub3:BubR1:Cdc20:APC complex. The recently discovered mitotic checkpoint factor 2 protein (MCF2) is a highly potent APC-inhibitor (module VI in Figure 4), yet the mechanism of binding to the APC and its regulation is still unknown (Braunstein et al., 2007; Eytan et al., 2008).

With the exception of MCF2, all complexes inhibiting APC rely on the presence of Cdc20:C-Mad2, which requires unattached kinetochores for adequately fast formation. We can categorize Cdc20:C-Mad2 complex (Dark gray inset in module III, <span style="color:red">Figure 4</span>) to be the "interface" connecting signaling from unattached kinetochores to APC inhibition.

Some studies (Ibrahim et al., 2008a; Ibrahim et al., 2008b; Ibrahim et al., 2009) indicate that for fast checkpoint deactivation after the last attachment destabilization of APC:inhibitor complexes might be required. However, the molecular basis of these effects is not yet understood, though it is reasonable to presume involvement of UbcH10, P31[comet], Usp44 and Dynein (Habu et al., 2002; Hagan et al., 2011; Reddy et al., 2007; Stegmeier et al., 2007; Vink et al., 2006; Wojcik et al., 2001; Xia et al., 2004; Yang et al., 2007).

### *Systems biology to the SAC mechanism*

Improper attachment of even a single kinetochore suffices to delay mitotic progression (Musacchio and Salmon, 2007). Though, it is still puzzling which molecular mechanisms can achieve reliable cell cycle arrest while remaining highly responsive. Whether sensing of tension is suitable and required for proper SAC response is not without debate (Khodjakov and Pines, 2010). To complicate the matter even further, the SAC must link the biomechanics of the mitotic spindle with a biochemical signal transduction network, thus posing an inherently spatial problem.

The pioneering work of Doncic et al. (Doncic et al., 2005) as well as Sear and Howard (Sear and Howard, 2006) analyzed simple spatial models of potential checkpoint mechanisms with focus on yeast or animals, respectively. To cope with the substantially larger animal cell volume, they propose two potentially complementary pathways, featuring a non-autocatalytic amplification step or active transport from the kinetochore towards the spindle pole. Ibrahim et al. (Ibrahim and Henze, 2014) show using spatial simulation that active transportation of Mad2 can greatly enhance Mad2:Cdc20 complex formation. In a very recent publication, Chen et al. (Chen and Liu, 2014) discussed an elaborate reaction-advection-diffusion-model of SAC in animal cells and emphasized the importance of streaming of SAC-components from attached kinetochores towards the centrosomes.

Ibrahim et al. (Ibrahim et al., 2008b) employ a straight-forward ODE-model of the "template model" with realistic parameters and find that neither autocatalytic amplification nor competitive inhibition of the template complexes can improve the model performance with respect to Cdc20-sequestration or release. The same authors developed comprehensive mechanistic models of the SAC to study the kinetics of MCC formation, and APC/C inhibition or activation (Ibrahim et al., 2008a; Ibrahim et al., 2009). A pivotal role for MCC in APC/C inhibition is predicted from the models.

The robustness of putative SAC signaling mechanisms to intrinsic noise has been studied (Doncic et al., 2006; Ibrahim et al., 2007). Doncic et al. constructed simplistic models which suggest that dimerization of the SAC key players, Mad2 and Cdc20, can serve as a low pass filter to reduce noise induced by fluctuations of the rate of Cdc20-degradation. Ibrahim et al. employ a significantly more elaborate SDE-model considering a discrete compartment for each kinetochore. The compartments are coupled by diffusion-like mass transfer, and it turns out that high diffusion rates can suppress the intrinsic noise of the kinetochore "micro-reactors".

Simonetta et al. (Simonetta et al., 2009) perform a thorough kinetic study driven by a detailed model of the SAC core mechanism. Lohel et al. (Lohel et al., 2009) build on this kinetic data and challenge the model proposed by Doncic et al.(Doncic et al., 2005) which assumes instantaneous activation and release of the inhibitor upon kinetochore contact. It turns out that this assumption is a critical over-simplification because accounting for realistic kinetochore-binding kinetics does significantly affect model performance.

Mistry et al. (Mistry et al., 2008) combines the conceptual model by Sear et al. (Sear and Howard, 2006) with an ODE model of the chromosome attachment state and does so provide a framework for integration with models accounting for the correction mechanism for improper chromosome attachment. Such a model could probably have been discussed by He et al. (He et al., 2011): they assume two antagonistic positive feedback loops linking chromosome tension with checkpoint activation and show that this topology makes cyclin degradation upon SAC silencing irreversible.

Finally, an interesting approach to deduce the kinetochore-related interaction network of the SAC was followed by Doncic et al. (Doncic et al., 2009) . They screen randomized network topologies for conformance with a suitably chosen set of *in vivo* deletion mutants and end up with a topology which is in good agreement with many experimental findings.

## 2.3. Placing the correct DNA into the right cell, the Spindle Position Checkpoint (SPOC)

Asymmetric cell division is crucial for the development of cellular diversity in multi-cellular organisms by generating mother and daughter cells with different fates (Chia et al., 2008; Januschke and Gonzalez, 2008; Segal and Bloom, 2001; Shapiro et al., 2002). The budding yeast S. cerevisiae is a unicellular organism which undergoes asymmetric cell division and has been widely used to study polarized cell growth and asymmetric cell division. Proliferating budding yeast cells establish cell polarity already in G1 phase upon commitment to a new cell cycle (Pruyne and Bretscher, 2000). To ensure proper chromosome segregation during asymmetric cell division in budding yeast, an additional mitotic checkpoint, the Spindle Position Checkpoint (SPOC) delays exit from mitosis until the mitotic spindle is correctly aligned along with the mother-bud axis (Bloecher et al., 2000; Pereira et al., 2000; Wang et al., 2000; Yeh et al., 1995) (for review see (Caydasi et al., 2010a; Merlini and Piatti, 2011)).

SPOC control involves phosphorylation events and alterations in the localization of proteins (Caydasi et al., 2010a). SPOC mechanism can be described by the regulation of two essential components, namely, Bfa1-Bub2 and Tem1.

### 2.3.1. Bfa1-Bub2 regulation

The SPOC governs the activity of the dimeric protein Bfa1-Bub2, which prevents initiation of late mitotic events. Bfa1-Bub2 is regulated through competing phosphorylation by the two kinases Cdc5 and Kin4. Phosphorylation by Cdc5 inactivates Bfa1-Bub2, while phosphorylation by Kin4 shields Bfa1 from the inactivating phosphorylation by Cdc5. Kin4 kinase activity in turn requires phosphorylation within its activation loop (T209 residue) by the bud neck kinase Elm1; however, this task seems to be accomplished independently of SPOC activation (Caydasi et al., 2010b). The protein phosphatase PP2A may contribute to SPOC signaling in two different ways, depending on whether it is bound to its regulatory subunit Cdc55 or Rts1. PP2A-Cdc55 has been implied in the removal of the hyperphosphorylation of Bfa1 by Cdc5 (Queralt et al., 2006), whereas PP2A-Rts1 promotes dephosphorylation of Kin4 and is required for association of Kin4 with the SPBs and the mother cell cortex (Chan and Amon, 2009).

Cdc5 localizes to both spindle pole bodies (SPB, centrosome equivalent in yeast) with similar amount and does not change in response to spindle misalignment (Maekawa et al., 2007). In contrast, upon spindle misalignment Bfa1-Bub2 and Tem1 localization changes from asymmetric to symmetric (Pereira et al., 2000). Similarly, SPOC activation leads in symmetric localization of Kin4 at the SPBs (D'Aquino et al., 2005; Pereira and Schiebel, 2005).

### 2.3.2. Tem1 activity control

The small GTPase Tem1 is the most upstream component of the mitotic exit network (MEN; (Amon, 2001; Jaspersen et al., 1998)). Tem1 is believed to be active in GTP bound state. Bfa1-Bub2 inhibits Tem1 activity by promoting GTP hydrolysis (Geymonat et al., 2002).

A central question is whether the GAP regulates Tem1 solely at the SPBs or everywhere (at the SPBs and also in the cytosol). Recently, a systems biology approach combining *in vivo* data with mathematical modeling (Caydasi et al., 2012) suggested that cytoplasmic regulation of Tem1 by Bfa1-Bub2 complex is critical for robust spindle position checkpoint arrest. The same study has shown the necessity of an additional mechanism for rapid Tem1-GTP vast recovery after spindle realignment.

Still, remaining to be answered is how the SPOC senses the spindle mis-orientation status and what additional control is required.

Taken together, both checkpoint SAC as well as SPOC are better understood when the checkpoints are active. However, missing mechanisms are required for checkpoint silencing controls.

### 2.3.3. Higher eukaryotes

In recent years much insight has been gained into the mechanisms and importance of spindle orientation in various organisms (for review see (Pereira and Yamashita, 2011; Siller and Doe, 2009)). Drosophila male GSCs have a SPOC-like mechanism, called the centrosome orientation checkpoint (COC), for delaying the onset of mitosis if centrosomes happen to be incorrectly oriented to the hub cell (Cheng et al., 2008).  The COC likely acts by altering the localization of Cyclin via the Par-1 kinase which is an AMPK-like kinase as Kin4 (Caydasi and Pereira, 2012; Cheng et al., 2008).

The main difference between SPOC and COC is the timing of cell cycle progression arrest. While COC delays the onset of mitosis, SPOC delays the onset of mitotic exit network control.

### *Why modeling SPOC*

Despite the low number of components, these components have various states and localizations upon which the interactions depend. Furthermore, these various states are difficult to analyze experimentally in living cells; however they are easily accessed in a model. For example; Tem1 activity whether it is GTP- or GDP bound, and Bfa1 phosphorylation by Kin4, by Cdc5, or not phosphorylated at all. Besides, answering pivotal questions like do the cytosolic pools play a role?

Using modeling and simulation also help to improve our understanding of the interplay of the components allowing us to understand what requirements the SPOC has to meet.

## 3. SAC and SPOC similarities

Although SAC and SPOC comprise different mitotic checkpoints, their mode of action has prominent similarities (Figure 5). Both pathways respond to a physical property of the spindle and rely on turnover of the inhibitor and activator at an organelle, broadcasting a 'WAIT' signal to the environment. The SAC integrates information about spindle attachment at the individual kinetochores which broadcast a nucleoplasmic (in budding yeast) 'WAIT'-signal unless proper attachment is established. The roles of Mad2 and Cdc20 are similar to the roles of Bfa1 and Tem1 in the SPOC, respectively (Figure 5A). Activity of Bfa1 is regulated at the SPBs which is broadcasted throughout the cytosol and inhibits the activator of the downstream pathway, Tem1 (Caydasi et al., 2012). Both, SAC and SPOC, must guarantee reliable broadcasting of the 'WAIT'-signal and simultaneously allow for its rapid removal when the checkpoint is satisfied (Figure 5B).

Mathematical modeling of SAC aided to exclude putative checkpoint architectures, which were not feasible when biophysical limitations (due to kinetochore-localization and interaction kinetics of central SAC components) were taken into account (Doncic et al., 2005; Ibrahim et al., 2008a; Ibrahim et al., 2008b; Sear and Howard, 2006). In addition, modeling of SAC helped to pinpoint advantages and problems of putative regulatory mechanisms (Ibrahim et al., 2008b; Sear and Howard, 2006; Simonetta et al., 2009). SPOC signaling is subject to similar biophysical limitations. In the same way successful application of modeling approaches contributed to the SAC field, mathematical modeling of SPOC has evaluated crucial hypothesis of cytoplasmic regulation of Tem1 by GAP complex in concordance with experimental results (Caydasi et al., 2012).

Such similarity plays a pivotal role to reveal many open questions. An example is that the components of the mitotic checkpoint complex (MCC), which forms as a result of SAC activation in animal cells, turn over at the kinetochores with residence half-lives of about 30 seconds or less (Maiato et al., 2004). This is comparable to the residency times of Bfa1 and Bub2 at the SPB when the SPOC is active. Kinetochores of budding yeast are comparable in size with the SPBs, thus numbers of the signaling molecules may be in the same order of magnitude. It would be interesting to see whether SAC signaling in yeast is constrained by the residency times in a way similar to the hyperphosphorylation of Bfa1 by Cdc5 in the SPOC.

Hyperphosphorylation of Bfa1 by Cdc5 may reduce the affinity of Bfa1 to Tem1 (Hu et al., 2001). However, this is not uncontroversial, because *in vitro* Bfa1 inhibits GTP-hydrolysis and GDP-GTP exchange of Tem1 in the absence of Bub2, and this inhibitory effect is not reduced if Bfa1 is hyperphosphorylated by Cdc5 (Geymonat et al., 2003). Hyperphosphorylated Bfa1 may instead have a reduced affinity to Bub2 (Geymonat et al., 2003). Combining that with our simulations suggests that phosphorylation of Bfa1 by Kin4 disables Bfa1 as a Cdc5-substrate, one might speculate that efficient formation of the Bfa1-Bub2 GAP-complex requires the SPBs, and that Cdc5 interrupts the association by hyperphosphorylation of Bfa1. In such a model, SPB-bound Kin4 would be required to ensure efficient association of Bfa1 with Bub2.

## 4. Conclusions and Perspectives

Both SAC and SPOC are intriguingly sensitive checkpoints. The underlying mechanisms communicate through biochemical signal transduction networks, which integrate, amplify, or attenuate signals in response to various inputs. Both checkpoints are intricate signaling cascades coupling the physiological state of the mitotic spindle with cell cycle progression. Their core mechanisms appear to be simple at first glance because only a handful of components interact, but significant complexity is added by dynamic localization to subcellular structures, multi-state-components, and the reversible formation of protein complexes thereof.

The close combination of experimental work with rigorous mathematical models was central to the success of physics in our modern world. Similarly, the systems biology approach exploits synergies from different disciplines to achieve a wholistic understanding of biological systems. Further experimentation combined with computational modeling will be needed to fill the gaps in our mechanistic understanding of mitosis. Such an integrative model could constitute a virtual wet-lab suitable for guiding further research to reveal the detailed mechanism behind mitotic regulation.

## Acknowledgements

We apologize to all those authors whose work could not be cited due to space constraints and the large number of studies performed in this field. We would like to thank Maiko Lohel for his thoughtful reading and commenting on the manuscript; Fouzia Ahmad for reading and correcting the manuscript.

## Conflict of interest

The author declares no conflict of interest.

**A  correct**

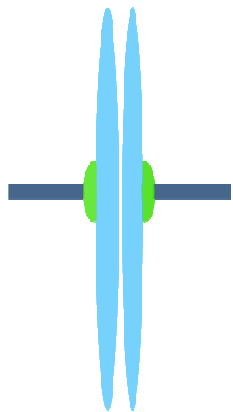

amphitelic

**B  incorrect**

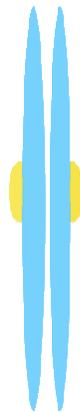

unattached

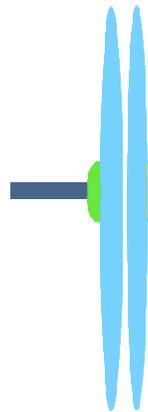

monotelic

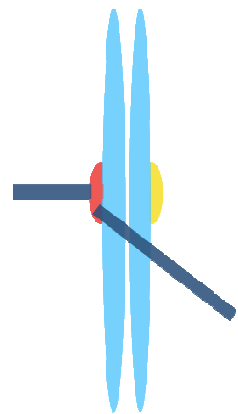

merotelic

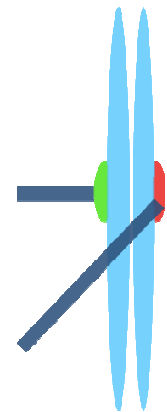

syntelic

**Figure 1**: **Scheme showing different anchoring between chromosomes and microtubules.** (A) Proper Bi-orientation chromosome segregation in anaphase requires all chromosomes to have amphitelic attachment, that is, both kinetochores of every chromosome must be attached to microtubules from opposite poles. (B) Before all chromosomes have established amphitelic attachment, chromosomes having no or erroneous attachments are frequent intermediates. Erroneous connections between microtubules and kinetochores spontaneously detach, facilitating proper re-attachment. Chromosomes showing merotelic attachment have a kinetochore which is simultaneously attached to microtubules from opposite poles. In contrast, chromosomes with syntelic attachment have both kinetochores connected to microtubules from the same pole.

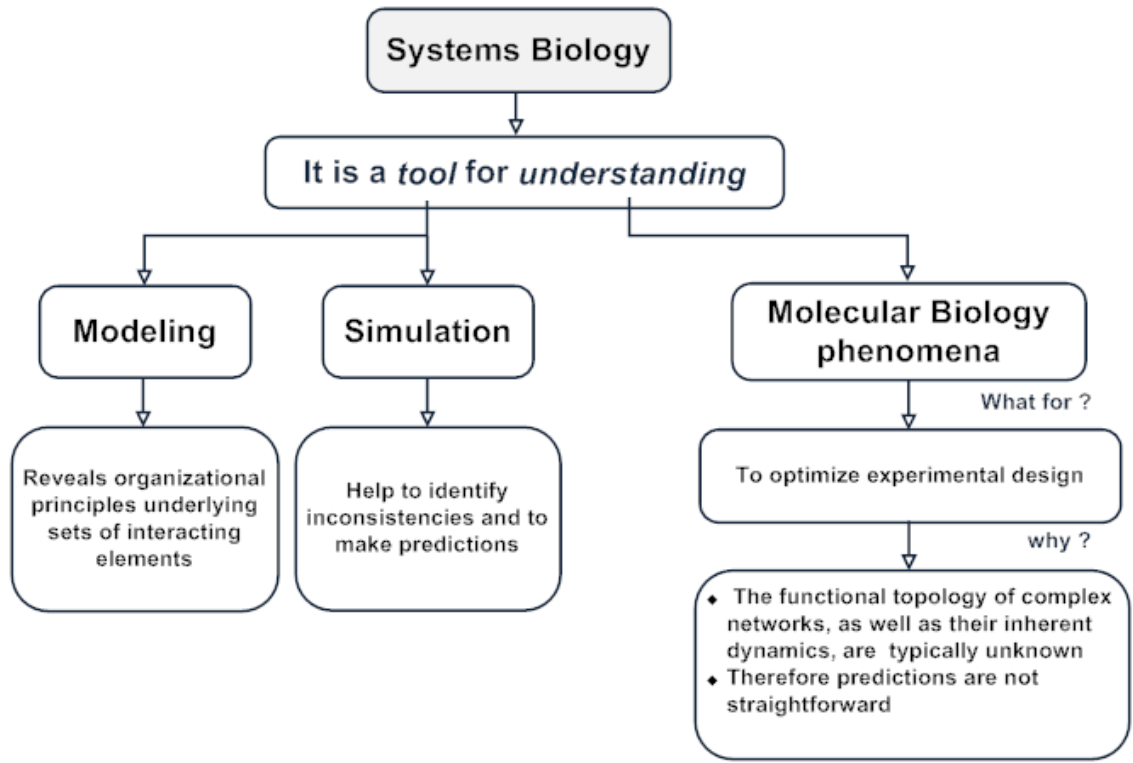

**Figure 2: Schematic flow chart illustrates the concept of "Systems Biology".** It can be define as a *tool* for *understanding* molecular biology phenomena through making prediction, designing experiments and testing hypotheses. It is based on *modeling* and *simulation*. *Modeling* reveals organizational principles underlying sets of interacting elements and *simulation* helps to identify inconsistencies and to make predictions.

## Systems Biology

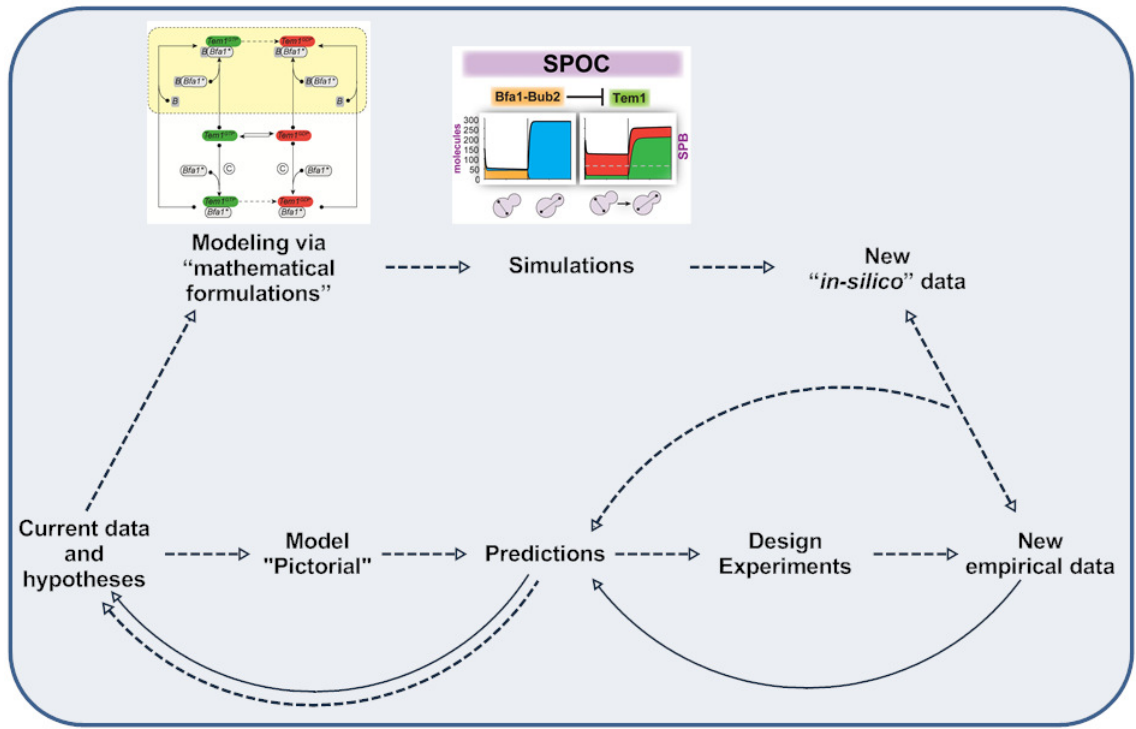

**Figure 3: A unifying framework shows a typical laboratory work scenario and a systems biology integrative approach.** In the first pathway cycle (solid arrows), we usually start in lab from data and hypotheses, subsequently we build a "pictorial model". It follows by suggesting hypotheses and predictions which are the base for designing new experiments. Then, we run these experiments for getting empirical data that are in turn correcting our predictions and hypotheses. In the second pathway cycle (dash arrows), overlaps with the experimental work, we build a quantitative mathematical model in which each element has a species meaning. Subsequently, we simulate our model to get *in-silico* data. These data together with the empirical data are interacted in order to correct our predictions, designing experiments and testing hypotheses. This framework which integrates both computational and experimental approach is called systems biology.

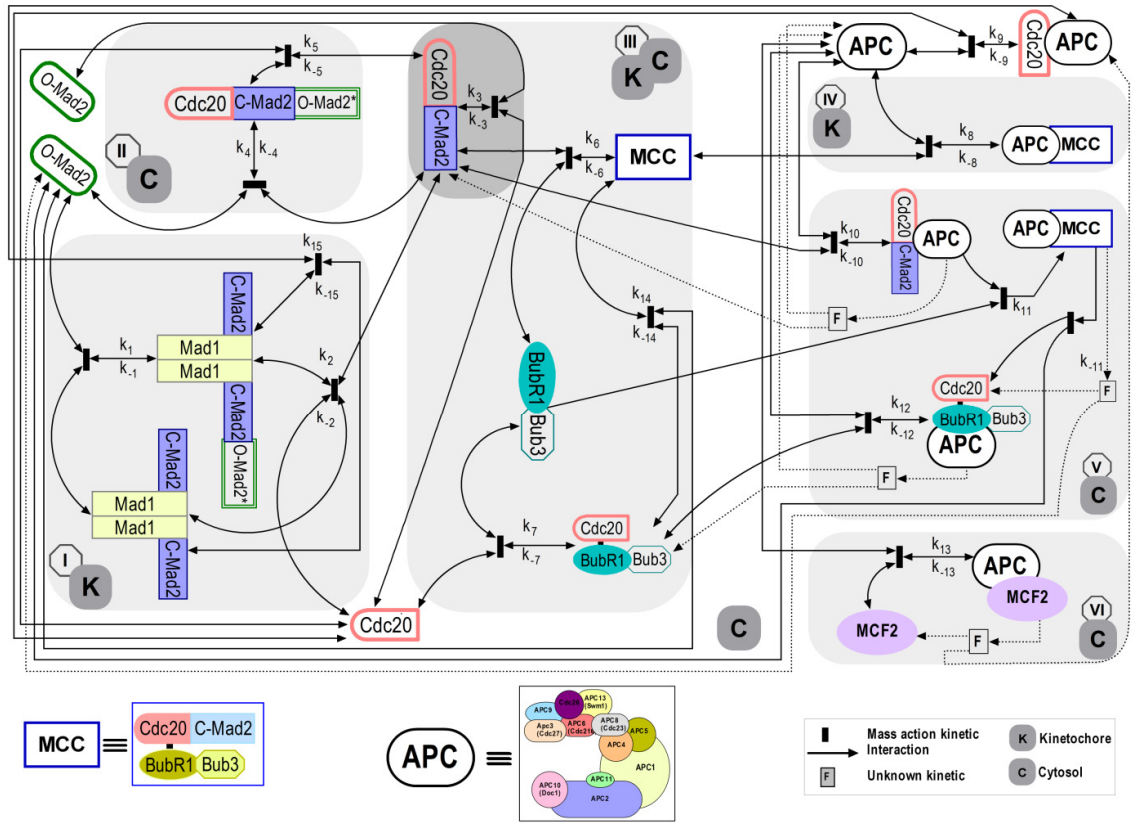

**Figure 4**: **Schematic illustrating the biochemical reaction network of SAC activation and maintenance.** Shown are all proteins/complexes, involved in the SAC mechanism, and the interactions between them. While the color boxes represent proteins, the arrows describe the interactions. The whole scheme is clustered in six parts or modules (I-VI). Modules I, III and IV are kinetochore dependent parts (K). Modules II, III, V and VI are the cytosolic parts (C). The basic "template model" is located in the modules I and III (kinetic parameters $k_1$, $k_2$, and $k_3$). The MCC formation is located in module III (kinetic parameters $k_4$). The APC regulation is located in modules IV, V, and VI (kinetic parameters $k_8$ to $k_{13}$).

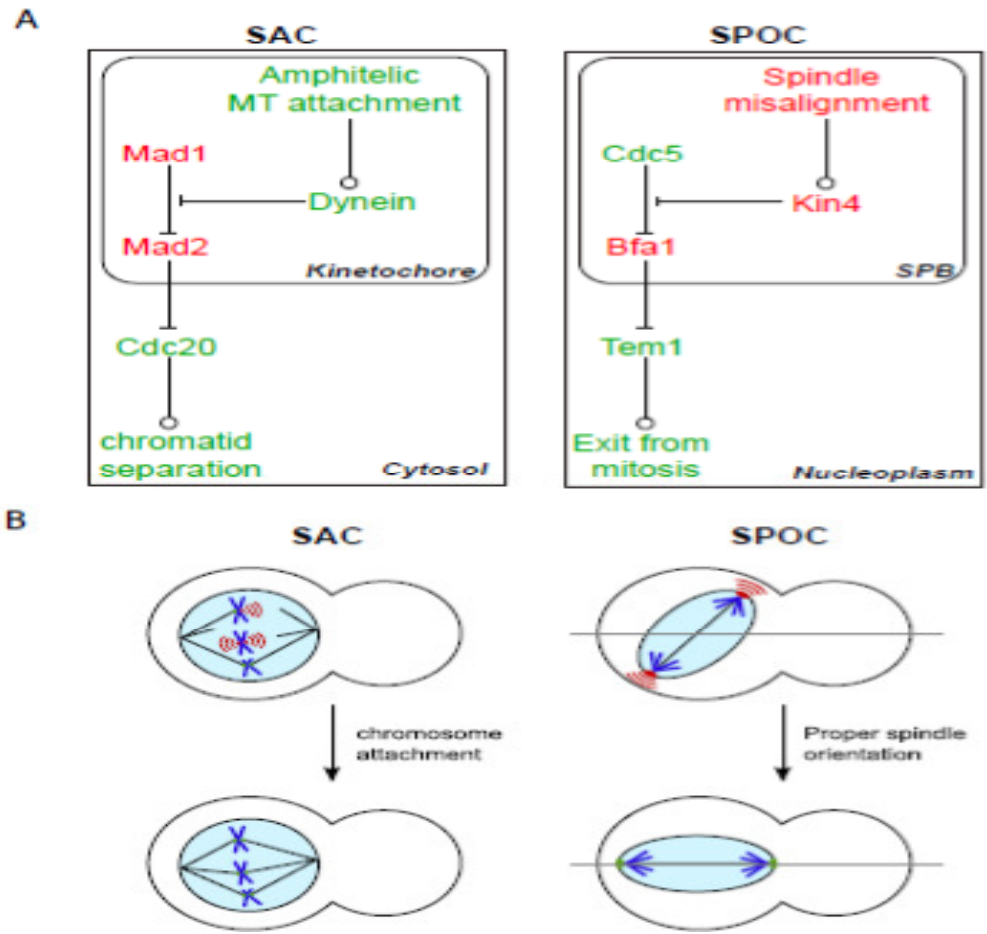

**Figure 5: SAC and SPOC similarities.** (A) SAC and SPOC pathways. (B) Mode of action in SAC (left) and SPOC (right). Central SAC components are recruited to unattached kinetochores and broadcast a 'WAIT'-signal to the nucleoplasm. Upon chromosome attachment, these components are stripped off the kinetochores and broadcasting ceases. Similarly, central SPOC components are localized at the SPBs and promote broadcasting of an inhibitory signal to the cytoplasm. When the spindle aligns correctly with the polarity axis, signaling from the SPBs is shut down.